\documentclass[showpacs,showkeys,prd,twocolumn]{revtex4-1}
\usepackage{amsmath}
\usepackage{graphicx}
\usepackage{amssymb}
\usepackage{natbib}

\newcommand{\pd}[2]{\frac{\partial #1}{\partial #2}}

\renewcommand{\vec}[1]{\mathbf{#1}}

\DeclareMathOperator{\erf}{erf}
\DeclareMathOperator{\Htheta}{\theta}
\DeclareMathOperator{\sgn}{sgn}

\begin{document}
\title{Self-consistent model of the electron-positron pair production in the collision of laser pulses}
\date{\today}
\author{Alexander Tarasenko}
\affiliation{Theoretical Physics Department, Belarusian State University, Belarus, Minsk, Nezavisimosti av., 4, 220030}
\email{tarasenk@tut.by}

\begin{abstract}
Pair production due to the Schwinger mechanism is modeled in a self-consistent approach taking into account the backreaction of the created particles on the field. The model uses the Boltzmann kinetic equation for the electrons and positrons and the Maxwell equations for the electromagnetic field, which are solved numerically in a certain approximation for the case of the linearly polarized plane wave collision. The results indicate three main stages of the collision process: the acceleration of particles, the deceleration of particles, and plasma oscillations. A linear equation describing the oscillations is obtained, the oscillation frequency is related to the frequency of the quasinormal mode of the plasma with the least damping. It is also shown that the oscillation frequency is of the order of the plasma frequency.
\end{abstract}



\maketitle

\section{Introduction}

The electron-positron pair production in strong electromagnetic field was predicted by Klein and Sauter (see Ref. \cite{RuffiniPhR2010} for a review), and modern laser facilities such as Extreme Light Infrastructure (ELI) \cite{ELI} and High Power laser Energy Research facility (HiPER) \cite{HiPER} offer a possibility to check these predictions. Potential intensities that can be reached at these facilities are up to $10^{26}$ W/cm${}^2$ \cite{HiPER}, which allows to observe physical processes taking place in an electric field close to the critical quantum field $E_c=m_e^2c^3/e\hbar$, which corresponds to the intensity of the order of $10^{29}$ W/cm${}^2$.

The phenomena taking place in a superstrong laser field have attracted much attention of researchers recently. Various processes have been investigated such as the electron-positron-photon cascade formation \cite{FedotovPhRL2010,BulanovPhRL2010,NerushPhR2011} and different quantum electrodynamical reactions including the nonlinear Breit-Wheeler and Bethe-Heitler effects \cite{HuPhRL2010}, photon-photon scattering \cite{LundstromPhRL2006}, and muon pair production \cite{KuchievPhRL2007}. \citet{FedotovPhRL2010} emphasize that if a cascade develops the influence of the particles on the initial electromagnetic field has to be taken into account. Such self-consistent model of cascade development in the field of colliding linearly polarized laser pulses has been recently proposed by \citet{NerushPhR2011}.

A self-consistent approach to pair production has been developed in a simpler model treating the electric field as uniform and considering only the Schwinger mechanism of pair production. Both the model based on the semiclassical quantum field theory \cite{KlugerPhRL1991} and the one based on the the kinetic description of the electron-positron plasma \cite{RuffiniPLB2003,VereshchaginPLA2007,Benedetti2011} predict plasma oscillations \cite{RuffiniPhR2010}. In a sufficiently large time when the pair production almost stops these oscillations can be interpreted as natural oscillations of the plasma. We may expect that the plasma formed during the collision of laser pulses also has natural oscillations that show themselves after the collision. The goal of the present paper is to model the evolution of the electron-positron plasma and the electromagnetic field in the collision of two linearly polarized laser pulses, taking into account the backreaction of the particles on the field.

The system (the electron-positron plasma and the field) will be modeled under the following assumptions: each of the pulses is a plane linearly polarized electromagnetic wave. The pairs are produced only due to the Schwinger mechanism, i. e. we neglect the radiation reaction force acting on each particle and thus the possibility of cascade formation. The electric field is of the order of or less than the critical one, and the characteristic scale of its variation is much less than the Compton wavelength of the electron $\lambda_e=\hbar/m_e c$, which implies that the particles are produced almost at rest and the pair production rate is described by the Schwunger formula. The electron-positron plasma is considered to be ideal, nondegenerate and collisionless. The plasma is treated in the ultrarelativistic approximation, i. e. we assume that in a negligible time after the creation the particle moves at the speed of light.

\section{Kinetic description of the electron-positron plasma}

Consider an electromagnetic field with the potentials $\varphi=0,\vec{A}=A_y(t,x)\vec{e}_y$, where $(x,y,z)$ are the Cartesian coordinates. The electric field strength $\vec{E}$ and the magnetic induction $\vec{B}$ are directed along $y$-axis and $z$-axis, respectively. The only nonzero components of these vectors are
\begin{equation}\label{EB vs A}
E_y=-\frac{1}{c}\pd{A_y}{t}, \quad B_z=\pd{A_y}{x}.
\end{equation}
Consider the fields for which $A_y(t,-x)=A_y(t,x)$ and thus $E_y(t,-x)=E_y(t,x)$ and $B_z(t,-x)=-B_z(t,x)$. Such field can model the collision of waves at the point $x=0$ where the magnetic field vanishes. We will consider the following initial field
\begin{equation}\label{A0}
\begin{split}
A^{(0)}_y(t,x)&=-\sqrt{\frac{\pi}{2}}E_0\sigma\Biggl(\erf{\left(\frac{c t-x-a}{\sqrt{2}\sigma}\right)}\\
&+\erf{\left(\frac{c t+x-a}{\sqrt{2}\sigma}\right)} \Biggr),
\end{split}
\end{equation}
where $\erf{x}$ is the error function
\begin{equation}\label{erf}
\erf{(x)}=\frac{2}{\sqrt{\pi}}\int\limits_0^x e^{-t^2}dt.
\end{equation}
Field (\ref{A0}) is a linear combination of two gaussian wavepackets with the amplitude (the maximum value of the electric field strength) $E_0$ and the characteristic size $\sigma$. The wavepackets are initially located at the distance $2a$ and collide at the moment $t=a/c$, the characteristic duration of the collision being $\sigma/c$. The maximum intensity of each wave is
\begin{equation}\label{W}
W=\frac{cE_0^2}{4\pi}=4.65\times10^{29}\left(\frac{E_0}{E_c}\right)^2\text{W/cm}{}^2.
\end{equation}
In order to ensure the applicability of the Schwinger formula for pair production and the ultrarelativistic approximation we assume that the following inequalities are satisfied:
\begin{equation}\label{E cons}
E_0\lesssim E_c, \quad \sigma\gg\lambda_e, \quad eE_0\sigma\gg m_e c^2.
\end{equation}

Denote $f_e(t,x,\vec{p})$ and $f_p(t,x,\vec{p})$ the distribution functions of electrons and positrons, respectively. Since the particles are created at rest they move in the $xy$ plane in field (\ref{EB vs A}). The Boltzmann kinetic equation for electrons reads
\begin{equation}\label{fe eq gen}
\begin{split}
\pd{f_e}{t}&+v_x\pd{f_e}{x}-\frac{e}{c}v_y B_z\pd{f_e}{p_x}\\
&-e\left(E_y-\frac{1}{c}v_x B_z\right)\pd{f_e}{p_y}=S(E_y,B_z)\delta(\vec{p}),
\end{split}
\end{equation}
where $\vec{v}$ is the electron velocity, $S(E,B)$ is the source describing the pair production due to the Schwinger mechanism \cite{Landau_QED}:
\begin{equation}\label{S}
\begin{split}
S(E,B)&=\frac{dN}{dt dV}=\Htheta{(E^2-B^2)}S(\sqrt{E^2-B^2}), \\
S(E)&=\frac{m_e^4 c^5}{4\pi^3\hbar^4}\left(\frac{E}{E_c}\right)^2\exp{\left(-\frac{\pi E_c}{E}\right)},
\end{split}
\end{equation}
and $\theta(x)$ is the Heaviside theta-function. Note that in general the pair production rate depends on two field invariants: $E^2-B^2$ and $\vec{E}\vec{B}$. For field (\ref{EB vs A}) the second invariant is identically zero, thus the pair production is described by the same expression as the one in the electric field $E'=\sqrt{E^2-B^2}$. The theta-function in expression (\ref{S}) prohibits pair production in a field that becomes a purely magnetic one in a certain reference frame. The source in equation (\ref{fe eq gen}) is proportional to the delta-function of the particle momentum which is a consequence of the assumption that the particles are produced at rest. It is valid if the characteristic energy obtained by a particle during the acceleration in the field is much greater than its initial energy, which is guaranteed by inequalities (\ref{E cons}).

The distribution function depends on five variables, which complicates the numerical solution of the kinetic equation. Therefore we replace the exact equation by an approximate one neglecting the terms proportional to $v_x$ and $B_z$ in equations (\ref{fe eq gen}). This approximation can be explained in the following way: the pair production rate in a subcritical field has a sharp maximum around the maximum point of the electric field, i. e. $x=0$. The magnetic field in this region is small due to the symmetry relation ($B_z(t,-x)=-B_z(t,x)$), and the velocity component $v_x$ is appears only due to the magnetic field and thus is also small as compared to the component $v_y$. We note that these conditions do not guarantee that the particle remains in the region with weak magnetic field all the time, and the particle really leaves this region. However, we will consider only the approximate model here assuming that it correctly predicts the qualitative properties of the system (see also Discussion section). The substitution
$f_e(t,x,\vec{p})=F_e(t,x,p_y)\delta(p_x)\delta(p_z)$ reduces the approximate equation to
\begin{equation}\label{Fe eq}
\pd{F_e}{t}-eE_y\pd{F_e}{p_y}=S(E_y,B_z)\delta(p_y).
\end{equation}
There is no need in solving the kinetic equation for the positrons since $f_p(t,x,p_x,p_y,p_z)=f_e(t,x,p_x,-p_y,p_z)$.

The Maxwell equations reduce to
\begin{equation}\label{A eq}
\pd{{}^2A_y}{x^2}-\frac{1}{c^2}\pd{{}^2A_y}{t^2}=-\frac{4\pi}{c}j_y,
\end{equation}
where the only nonzero component of the conduction current is
\begin{equation}\label{jy}
j_y(t,x)=-2ec\int\limits_{-\infty}^{\infty}\sgn{p_y}F_e(t,x,p_y)dp_y,
\end{equation}
where $\sgn{x}$ is the signum function. There is no polarization current on the right side of equation (\ref{A eq}) since it is identically zero for source (\ref{fe eq gen}) in the ultrarelativistic approximation.

We solved equation system (\ref{Fe eq})-(\ref{jy}) numerically with the following initial and boundary conditions
\begin{equation}\label{inc}
\begin{split}
&A_y(0,x)=A^{(0)}_y(0,x), \quad \pd{}{t}A_y(0,x)=\pd{}{t}A^{(0)}_y(0,x),\\
&A_y(t,\pm X_0)=0, \quad F_e(0,x,p_y)=0,\\
&F_e(t,\pm X_0,p_y)=0, \quad F_e(t,x,\pm p_{max})=0,
\end{split}
\end{equation}
where $X_0$ and $p_{max}$ are sufficiently large numbers determining the boundary of the integration domain. We fixed the parameters $a=5\sigma$ and $\sigma=10^{-4}$ cm (which corresponds to the laser wavelength of the order of $1\mu$m), and varied $E_0$ from $0.065E_c$ to $0.23E_c$. A system of units with $\sigma=c=E_c=1$ was used in calculations, so that
\begin{equation}\label{pars}
\quad \lambda_e=3.86\times 10^{-7}, \quad e=\alpha E_c\lambda_e^2=1.09\times10^{-15},
\end{equation}
where $\alpha=e^2/\hbar c$ is the fine structure constant. The corresponding intensities of the initial waves (\ref{W}) varied from $1.96\times10^{27}\text{W/cm}{}^2$ to $2.46\times10^{28}\text{W/cm}{}^2$. The first value is approximately the pair production threshold: for $E_0=E_{min}=0.0644E_c$ the pair production rate is $S(E_{min})=c\sigma^{-4}$. The second value, as the numerical solution showed, corresponds to a regime when the backreaction of the particles significantly changes the initial field. Note that the adiabaticity parameter $\eta=\sigma E_0/\lambda_e E_c$ \cite{VereshchaginPLA2007} is large in the range of parameters $E_0$ and $\sigma$ used in calculations. The condition $\eta\gg1$ is necessary to use the Schwinger formula for the pair production rate.

Equation system (\ref{Fe eq})-(\ref{jy}) was solved in the region $0<t<T_0$, $-X_0<x<X_0$, $-p_{max}<p_y<p_{max}$, where $T_0=X_0=15$, $p_{max}=0.5$. The finite difference method was used, the grid having $N_x=251$ points in the $x$-direction and $N_p=401$ point in the $p_y$-direction. Due to strong dependance of the pair production rate on the field strength it was sufficient to solve equation (\ref{Fe eq}) in the range $-11\Delta x\leqslant x\leqslant11\Delta x$, where $\Delta x=0.12$ is the grid spacing. Since the source in equation (\ref{Fe eq}) is singular (proportional to the delta-function) the equation was not solved directly. The solution was expressed as
\begin{equation}\label{F real sol}
\begin{split}
F_e(t,x,p_y)&=\Htheta{(-p_y)}F^{(-)}_e(t,x,p_y) \\
&+\Htheta{(p_y)}F^{(+)}_e(t,x,p_y),
\end{split}
\end{equation}
and both functions $F^{(\pm)}_e(t,x,p_y)$ were obtained by solving the corresponding homogeneous equation. Equation (\ref{Fe eq}) itself was used for solution joining:
\begin{equation}\label{Fe joining}
F^{(-)}_e(t,x,0)-F^{(+)}_e(t,x,0)=\frac{1}{eE_y}S(E_y,B_z).
\end{equation}

\begin{figure}[ht]
\begin{tabular}{c}
\includegraphics{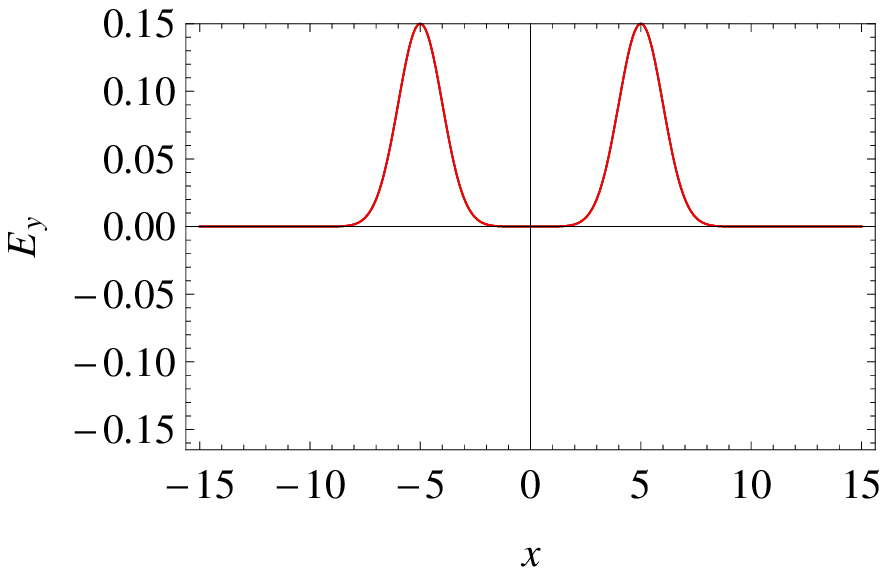} \\
(a) \\
\includegraphics{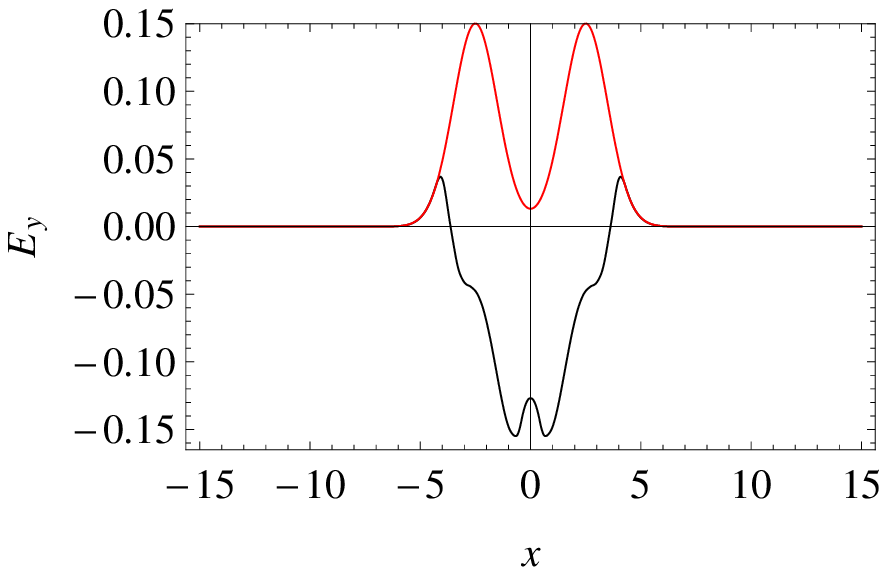}\\
(b) \\
\includegraphics{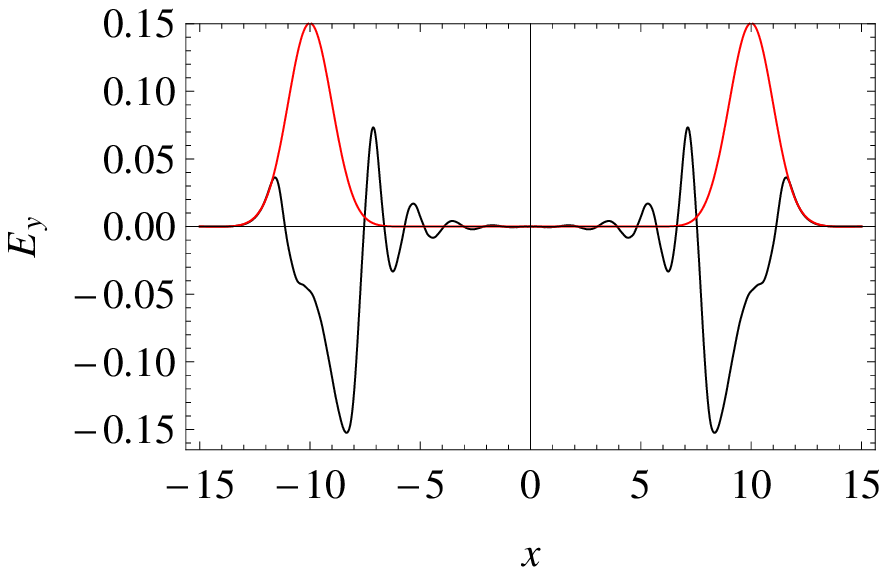}\\
(c) \\
\end{tabular}
\caption{(color online) Electric field $E_y(t,x)$ as a function of $x$ for $t=0$ (a), $t=0.5 T_0$ (b), and $t=T_0$ (c) for the solution with $E_0=0.15$ (black) and for initial field (\ref{A0}) (red).}
\label{Es fig}
\end{figure}

The electric field $E_y$ at different moments of time for a specific solution with $E_0=0.15$ is shown in Fig. \ref{Es fig}. It follows from the figure that the backreaction of the particles on the field significantly changes the latter. It is also evident that at some points the electric field has the opposite direction to the direction of the initial field (\ref{A0}).

\begin{figure*}[ht]
\begin{tabular}{cc}
\includegraphics{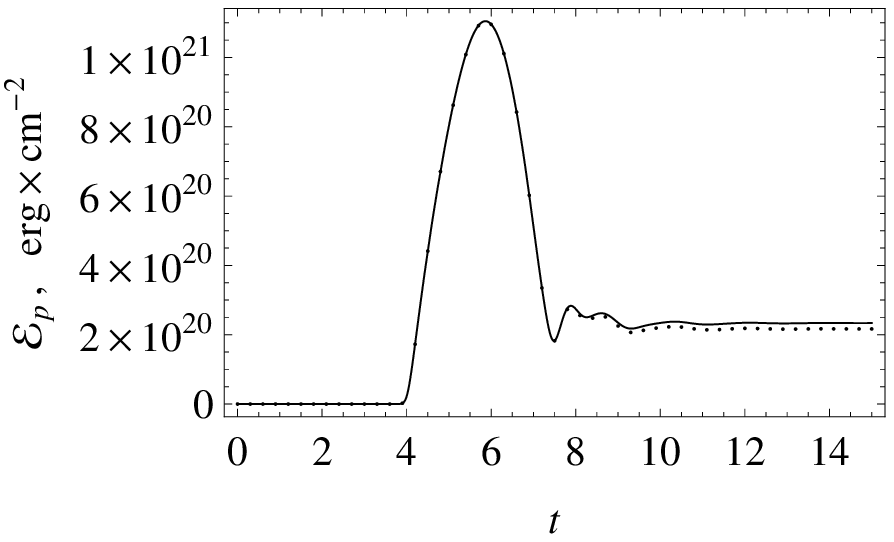} & \includegraphics{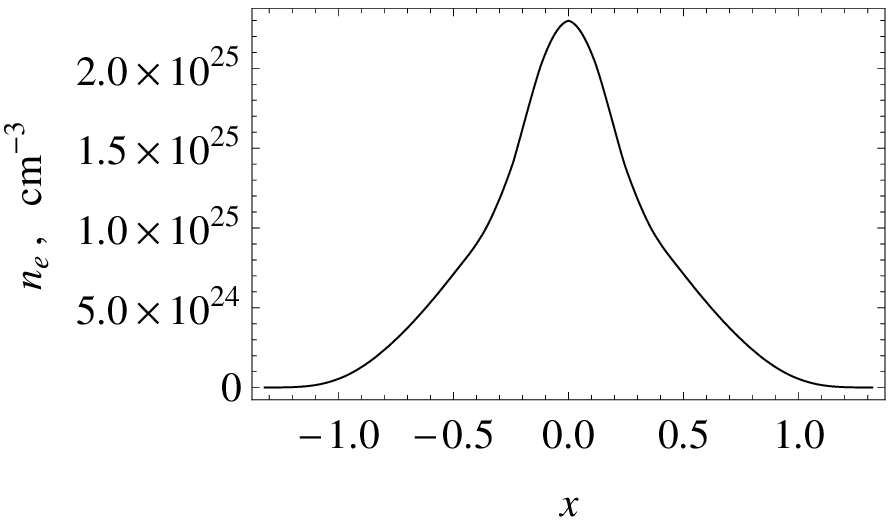}\\
(a) & (b) \\
\end{tabular}
\caption{(a) Energy of particles as a function of time for the solution with $E_0=0.15$ obtained from equation (\ref{Enp}) (solid line) and as ${\cal E}_f(0)-{\cal E}_f(t)$ using equation (\ref{Enf}) (shown by points). The coincidence of both plots implies the conservation of the total energy. The latter is equal to ${\cal E}=1.24\times10^{21}\text{erg}\times \text{cm}^{-2}.$ (b) The concentration of electrons as a function of $x$ for the same solution.}
\label{Ene fig}
\end{figure*}

Equation system (\ref{Fe eq})-(\ref{jy}) is self-consistent, which leads to the conservation of the total energy ${\cal E}={\cal E}_f+{\cal E}_p$, where
\begin{equation}\label{Enf}
{\cal E}_f=\frac{1}{8\pi}\int\limits_{-\infty}^{\infty}(E_y^2+B_z^2) dx\\
\end{equation}
is the field energy and
\begin{equation}\label{Enp}
{\cal E}_p=2\int\limits_{-\infty}^{\infty}\int\limits_{-\infty}^{\infty}|p_y|F_e dp_y dx
\end{equation}
is the energy of particles, both related to a unit area in the $yz$ plane.
Fig. \ref{Ene fig}a shows the energy of particles as a function of time. It follows from the figure that at first a significant part of the total energy (about $90\%$ for $E_0>0.1$) is transmitted to the particles. Then the electric field changes its direction and a large part of the energy of particles is returned back to the field. For $E_0>0.1$ the particles possess from $10\%$ to $40\%$ of the total energy in the final state, the percentage increasing with $E_0$. The calculation shows that the total energy is conserved with the maximum deviation not exceeding $2\%$ for all values of $E_0$.

The concentration of electrons (or positrons) is given by
\begin{equation}\label{ne}
n_e=n_p=\int\limits_{-\infty}^{\infty}F_e dp_y.
\end{equation}
The numerically calculated electron concentration is presented in Fig. \ref{Ene fig}b. The figure shows that the pairs are produced in a region with a characteristic size $\sigma$ centered at the point $x=0$. Note that plasma with a characteristic Lorenz factor of particles equal to $10^5$ and a concentration of the order of $10^{25}$cm${}^{-3}$ can be considered as ideal and nondegenerate, which was initially assumed.

\begin{figure}[ht]
\includegraphics{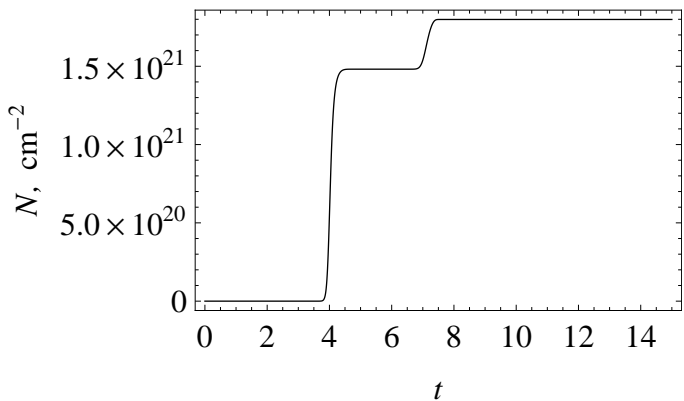}
\caption{Number of pairs $N$ as a function of time for the solution with $E_0=0.15$.}
\label{N fig}
\end{figure}

The total number of pairs per unit area can be written as
\begin{equation}\label{N}
N=\int\limits_{-\infty}^{\infty} n_e dx=\int\limits_{-\infty}^{\infty}\int\limits_{-\infty}^{\infty}F_e dp_y dx.
\end{equation}
The evolution of the total number of pairs is shown in Fig. \ref{N fig}. It follows from the figure that pairs are produced in significant quantities only at certain moments of time which, as the analysis shows, approximately coincide with the maximum points of the electric field strength $|E_y|$ at the origin $x=0$.

We also found the number and energy of pairs in a non-self-consistent model the basic equations of which consist of equation (\ref{Fe eq}) and (\ref{A eq}) without the source on the right side. The self-consistent model always predicts smaller values both for the number of particles and their energy, the difference in predictions increasing with $E_0$: for $E_0=0.065$ the ratio of the numbers and energies of particles in the final state in self-consistent and non-self-consistent models are $0.99$ and $0.94$, respectively, and for $E_0=0.23$ these ratios are $6.0\times10^{-7}$ and $8.3\times10^{-8}$.

\section{Linearized model}

Numerical solution gives the evidence of plasma oscillations after the collision, which are present in Fig. \ref{Es fig}c. When the magnitude of the electric field is sufficiently small we expect the oscillations to be linear. In this case the quantity $A_y(t,x_1)$ as a function of time at a fixed point $x=x_1$ has to be of the following form:
\begin{equation}\label{A gen lin}
A_y(t,x_1)=\tilde{A}_y+A_0e^{-\lambda(t-t_a)}\sin{(\omega_0(t-t_a))},
\end{equation}
where the parameters $\tilde{A}_y,A_0,\omega_0,t_a,\lambda$ do not depend on time. The potential (\ref{A gen lin}) possesses two properties: firstly the moments of time $t_n$ form an arithmetical progression, where $t_n$ are the roots of equation $E_y(t_n,x_1)=0$ enumerated in ascending order; secondly the quantities $(A_y(t_{n+1},x_1)-A_y(t_{n},x_1))$ form a geometric progression. We chose $x_1=0$ and found that these two properties hold for all values of $E_0$ in the range $0.065\leq E_0\leq0.23$ if we omit several first moments $t_n$ at which the oscillations are still nonlinear. The common difference and the common ratio of the progressions allow to calculate the oscillation frequency $\omega_0$ and the parameter $\lambda$. Particularly, for $E_0=0.15$
\begin{equation}\label{freq E0 8}
\omega_0=3.49, \quad \lambda=0.785.
\end{equation}

The linear regime of oscillations is described by linear equations that follow from equation system (\ref{Fe eq})-(\ref{jy}). In this regime the pair production is negligible, therefore equation (\ref{Fe eq}) has a solution
\begin{equation}\label{Psi}
F_e(t,x,p_y)=\Psi\left(x,p_y-\frac{e}{c}A_y(t,x)\right),
\end{equation}
where $\Psi(x,p_y)$ is an arbitrary function. Equation (\ref{Psi}) implies that the plot of $F_e$ versus $p_y$ for a fixed value of $x$ shifts along $p_y$-axis with time, its form remaining constant. Introduce the following notations:
\begin{equation}\label{delta A}
\begin{split}
A_y(t,x)&=\tilde{A}_y(x)+\delta A_y(t,x), \\
\tilde{A}_y(x)&=A_y(\infty,x).
\end{split}
\end{equation}
In the linear in $\delta A_y$ approximation
\begin{equation}\label{j lin}
j_y(t,x)=\tilde{j}_y(x)-4e^2\delta A_y(t,x)\Psi\left(x,-\frac{e}{c}\tilde{A}_y(x)\right),
\end{equation}
where $\tilde{j}_y(x)=j_y(\infty,x)$. Substitution of equation (\ref{j lin}) into (\ref{A eq}) yields
\begin{equation}\label{delta A eq}
\left(\pd{{}^2}{x^2}-\frac{1}{c^2}\pd{{}^2}{t^2}-V(x)\right)\delta A_y=0,
\end{equation}
\begin{equation}\label{V}
V(x)=\frac{16\pi e^2}{c}F_e(\infty,x,0),
\end{equation}
where relation (\ref{Psi}) and the fact that $\delta A_y=0$ is a solution of the linearized equation were used.

Equation (\ref{delta A eq}) is a wave equation with the potential $V(x)$. Numerical solution of equation system (\ref{Fe eq})-(\ref{jy}) together with equation (\ref{V}) allows to find the potential for arbitrary values of the parameters. The analysis shows that the potential is always positive, symmetric ($V(-x)=V(x)$) and tends to zero at the infinity. For $E_0<E_1\approx0.15$ the potential has a single maximum at the point $x=0$. If $E_0>E_1$ then the maxima of the potential are located symmetrically with respect to the origin and $x=0$ is the minimum point. Equation (\ref{delta A eq}) describes the penetration of a wave through a potential barrier. The asymptotic form of the function $\delta A_y$ for large $t$ can be described in terms of the quasinormal modes of oscillations \cite{KokkotasLivRev1999}, which are analogous to the quasidiscrete states of a quantum system. For sufficiently large $t$ the quantity $\delta A_y$ is approximately proportional to the quasinormal mode with the least damping. The quasinormal mode is a solution of equation (\ref{delta A eq}) of the form $\delta A_y=\psi(x)\exp{(-i\omega t)}$, where $\omega=\omega_0-i\lambda$ (the parameters $\omega_0$ and $\lambda$ have the same meaning as in expression (\ref{A gen lin})). The function $\psi$ satisfies the following equation
\begin{equation}\label{psi eq}
\psi''(x)+\left(\frac{\omega^2}{c^2}-V(x)\right)\psi(x)=0,
\end{equation}
where a prime denotes differentiation of a function with respect to its argument, together with the boundary conditions
\begin{equation}\label{psi inc 1}
\begin{split}
\psi(-X_1)&=\exp{\left(-i\frac{\omega}{c} X_1\right)}, \\
\psi'(-X_1)&=-i\frac{\omega}{c}\exp{\left(-i\frac{\omega}{c} X_1\right)},
\end{split}
\end{equation}
\begin{equation}\label{psi inc 2}
\psi'(X_1)-i\frac{\omega}{c} \psi(X_1)=0,
\end{equation}
where $X_1$ is sufficiently large so that the inequality $V(\pm X_1)\ll\omega^2/c^2$ holds (for an exact quasinormal mode $X_1=\infty$). We found the quasinormal mode frequency numerically integrating equation (\ref{psi eq}) with initial conditions (\ref{psi inc 1}) for a fixed value of $\omega$, and then solving equation (\ref{psi inc 2}) as an algebraic one. The initial condition required for the numerical solution of equation (\ref{psi inc 2}) was $\omega^{(0)}=c\sqrt{V_{max}}$, where $V_{max}$ is the maximum value of the potential (the first quasidiscrete state is located near the maximum point of the potential energy, see also \cite{KokkotasLivRev1999}).

\begin{figure*}[ht]
\begin{tabular}{cc}
\includegraphics{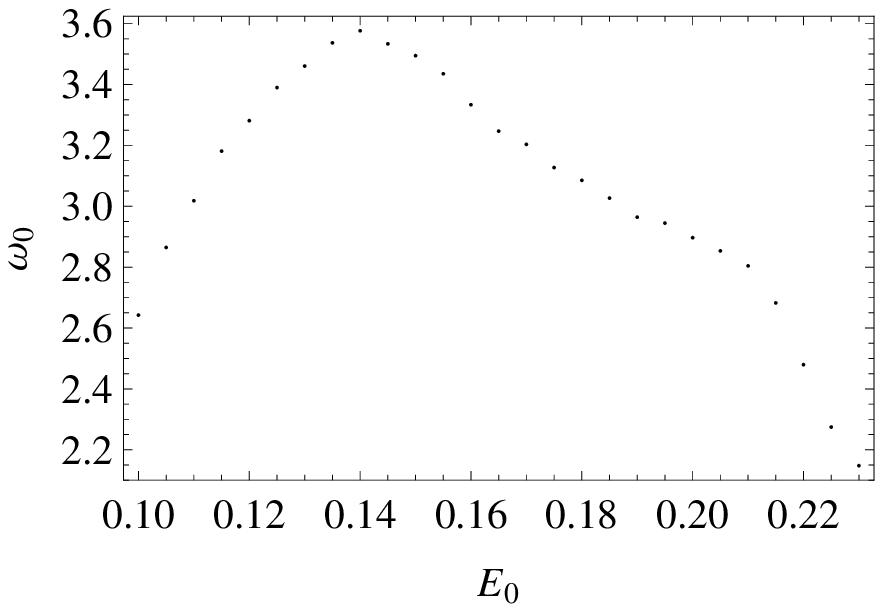} & \includegraphics{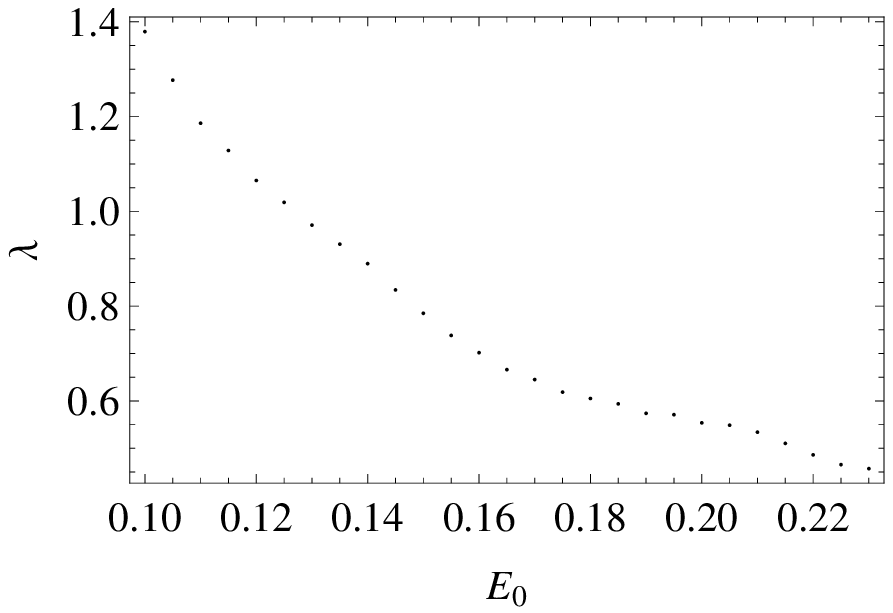}\\
(a) & (b) \\
\end{tabular}
\caption{Oscillation frequency $\omega_0$ (a) and parameter $\lambda$ (b) as a function of $E_0$.}
\label{ol pl}
\end{figure*}

The calculated parameters $\omega_0$ and $\lambda$ are presented in Fig. \ref{ol pl}. The data shown in the figure were obtained similarly to the values (\ref{freq E0 8}), i. e. by solving nonlinear equation system (\ref{Fe eq})-(\ref{jy}). The values of $\omega_0$ and $\lambda$ were calculated only for $E_0\geq0.1$ since in the opposite case we could not observe enough oscillations. The results of the quasinormal mode calculation lead to the same values of the parameters with a maximum deviation of less than $2\%$ both for $\omega_0$ and $\lambda$. The plots (Fig. \ref{ol pl}) show a nontrivial dependance of the parameters on the field strength, particularly, no asymptotic behavior both for large and for small values of $E_0$ can be revealed. The same conclusion is expected about the dependance of $\omega_0$ and $\lambda$ on $\sigma$.

The oscillation frequency $\omega_0$ can be estimated for $E_0<E_1$ given the characteristic Lorenz factor $\gamma$ and the concentration of electrons $n_e$. Using the characteristic value of the momentum $m_e c\gamma$ and equation (\ref{ne}) we obtain: $n_e(t,x)\sim F_e(t,x,0)m_e c \gamma$. The approximation $\omega_0\approx c\sqrt{V(0)}$ (for $E_0<E_1$ the maximum value of the potential is $V(0)$) and formula (\ref{V}) lead to the estimate
\begin{equation}\label{omega0 est}
\omega_0\sim\sqrt{\frac{16\pi e^2n_e}{m_e\gamma}}.
\end{equation}
\citet{Benedetti2011} introduce the notion of the relativistic plasma frequency $\omega_p$, which is $\sqrt{2}$ times less than the right side of equation (\ref{omega0 est}) (it should be noted that in \cite{Benedetti2011} all electrons are assumed to have the same Lorenz factor, and $\gamma$ is understood as its value). Estimate (\ref{omega0 est}) shows that the oscillation frequency is of the order of the plasma frequency. The parameter $\lambda$ is related to the quantity $V''(0)$ \cite{KokkotasLivRev1999} and thus is more difficult to estimate.

\section{Discussion}

The results of the modeling suggest the following scenario of the laser pulse collision. Pair production begins approximately at the moment of collision and occurs in a small region the size of which is of the order of the laser wavelength. The created particles are accelerated to ultrarelativistic energies by the electric field. The particles produce electric current which causes the electric field to change its direction, and the energy of particles is returned back to the field. This process can repeat several times. After the magnitude of the field becomes sufficiently small the particle production almost stops, the field energy remains approximately constant, and the plasma reaches the equilibrium state through oscillations.

The advantage of the considered model is that the kinetic equation for the electrons is reduced to an equation in three variables, and even this equation has to be solved in a small region close to the origin ($|x|\lesssim\sigma$). However, this allows to obtain only qualitative but not quantitative description of the system. The direct solution of the equation of motion shows that the electron leaves the region where the magnetic field is weak in time (the escape time is of the order of $\sigma/c$). This leads to the expansion of the plasma cloud, which was also obtained by \citet{NerushPhR2011}. Moreover, the radiation losses of a particle become important since the particles are accelerated to ultrarelativistic energies in subcritical fields (in agreement with \cite{VereshchaginPLA2007}). A self-consistent description of radiation requires taking into account the existence of photons and the possibility of pair production by them. The processes of radiation and pair production by a photon result in the possibility of the electron-positron-photon cascade formation. Nevertheless, we expect that these two effects do not change the general conclusions drawn above, particularly about the existence of plasma oscillations. The expansion of the plasma is expected to reduce the oscillations: the oscillation frequency has to decrease due to the decrease in particle concentration (\ref{omega0 est}), and the parameter $\lambda$ has to increase since the plasma cloud loses energy not only due to the radiation but also due to the particle loss. The cascade formation is expected to increase the oscillation frequency as it increases the particle concentration (it is more difficult to draw a conclusion about the parameter $\lambda$: it has to decrease if the maximum of the potential $V(0)$ becomes less sharp and thus $V''(0)$ decreases, but it is not clear whether such effect takes place).

A model that is free of these two approximations but does not take into account the Schwinger mechanism of pair production was developed by \citet{NerushPhR2011}. Paper \cite{NerushPhR2011} considers the cascade formation in the collision of linearly polarized laser pulses having finite width in $x$ and $y$-directions. The model is based on the combination of the particle-in-cell (PIC) and Monte-Carlo algorithms, and is self-consistent. However, neither energy transmission from the particles to the field nor plasma oscillations were obtained as the result of modeling. We can explain it by the fact that the numerical solution was stopped when the energy of all particles was roughly $40\%$ of the total energy. The results of the present paper suggest that the energy transmission to the field starts when the energy of particles becomes approximately $90\%$ of the total energy, and the oscillations start even later.

Plasma oscillations were obtained in the uniform plasma model \cite{VereshchaginPLA2007,Benedetti2011}. The plasma oscillations in this model in the weak field limit are interpreted as the longitudinal Langmuir waves having an infinite wavelength and thus the frequency equal to the plasma one. The electromagnetic radiation is impossible in this model thus the field energy periodically
vanishes being totally transmitted to the particles, and the oscillation damping becomes negligible for large $t$. In the given paper the geometry of the field is different thus several conclusions change: the field energy is not totally transmitted to the particles, the percentage of the transmitted energy depending on the intensity of the initial waves; the oscillations are damped due to the electromagnetic radiation. However, both models predict the same estimates of the parameters: the oscillation frequency is of the order of the plasma frequency \cite{Benedetti2011}, the characteristic Lorenz factor of the particles is of the order of $10^5$ for $E_0\sim0.15$ \cite{VereshchaginPLA2007}.

\section{Conclusions}

In the given work a self-consistent model of pair production in the collision of two linearly polarized plane electromagnetic waves is presented. The general scenario of collision is described; it consists of three stages: the acceleration of particles, the deceleration of particles, and plasma oscillations. The linearized model describing the plasma oscillations is obtained from the general nonlinear one. The parameters of oscillations are related to the complex frequency of the quasinormal mode with the least damping. Estimate of the oscillation frequency (\ref{omega0 est}) is obtained. The effects that are not taken into account in the model are considered, and their role is discussed. The results of this paper are compared with the results obtained using different models.

\begin{acknowledgements}
The author would like to thank G. Vereshchagin for useful advice and discussion.
\end{acknowledgements}

\bibliographystyle{apsrev4-1}
\bibliography{pair_prod_biblio}

\begin{thebibliography}{15}%
\makeatletter
\providecommand \@ifxundefined [1]{%
 \@ifx{#1\undefined}
}%
\providecommand \@ifnum [1]{%
 \ifnum #1\expandafter \@firstoftwo
 \else \expandafter \@secondoftwo
 \fi
}%
\providecommand \@ifx [1]{%
 \ifx #1\expandafter \@firstoftwo
 \else \expandafter \@secondoftwo
 \fi
}%
\providecommand \natexlab [1]{#1}%
\providecommand \enquote  [1]{``#1''}%
\providecommand \bibnamefont  [1]{#1}%
\providecommand \bibfnamefont [1]{#1}%
\providecommand \citenamefont [1]{#1}%
\providecommand \href@noop [0]{\@secondoftwo}%
\providecommand \href [0]{\begingroup \@sanitize@url \@href}%
\providecommand \@href[1]{\@@startlink{#1}\@@href}%
\providecommand \@@href[1]{\endgroup#1\@@endlink}%
\providecommand \@sanitize@url [0]{\catcode `\\12\catcode `\$12\catcode
  `\&12\catcode `\#12\catcode `\^12\catcode `\_12\catcode `\%12\relax}%
\providecommand \@@startlink[1]{}%
\providecommand \@@endlink[0]{}%
\providecommand \url  [0]{\begingroup\@sanitize@url \@url }%
\providecommand \@url [1]{\endgroup\@href {#1}{\urlprefix }}%
\providecommand \urlprefix  [0]{URL }%
\providecommand \Eprint [0]{\href }%
\providecommand \doibase [0]{http://dx.doi.org/}%
\providecommand \selectlanguage [0]{\@gobble}%
\providecommand \bibinfo  [0]{\@secondoftwo}%
\providecommand \bibfield  [0]{\@secondoftwo}%
\providecommand \translation [1]{[#1]}%
\providecommand \BibitemOpen [0]{}%
\providecommand \bibitemStop [0]{}%
\providecommand \bibitemNoStop [0]{.\EOS\space}%
\providecommand \EOS [0]{\spacefactor3000\relax}%
\providecommand \BibitemShut  [1]{\csname bibitem#1\endcsname}%
\let\auto@bib@innerbib\@empty
\bibitem [{\citenamefont {Ruffini}\ \emph {et~al.}(2010)\citenamefont
  {Ruffini}, \citenamefont {Vereshchagin},\ and\ \citenamefont
  {Xue}}]{RuffiniPhR2010}%
  \BibitemOpen
  \bibfield  {author} {\bibinfo {author} {\bibfnamefont {R.}~\bibnamefont
  {Ruffini}}, \bibinfo {author} {\bibfnamefont {G.}~\bibnamefont
  {Vereshchagin}}, \ and\ \bibinfo {author} {\bibfnamefont {S.-S.}\
  \bibnamefont {Xue}},\ }\href {\doibase DOI: 10.1016/j.physrep.2009.10.004}
  {\bibfield  {journal} {\bibinfo  {journal} {Physics Reports}\ }\textbf
  {\bibinfo {volume} {487}},\ \bibinfo {pages} {1 } (\bibinfo {year}
  {2010})}\BibitemShut {NoStop}%
\bibitem [{ELI()}]{ELI}%
  \BibitemOpen
  \href@noop {} {}\bibinfo {howpublished}
  {www.extreme-light-infrastructure.eu}\BibitemShut {NoStop}%
\bibitem [{HiP()}]{HiPER}%
  \BibitemOpen
  \href@noop {} {}\bibinfo {howpublished} {www.hiper-laser.org}\BibitemShut
  {NoStop}%
\bibitem [{\citenamefont {Fedotov}\ \emph {et~al.}(2010)\citenamefont
  {Fedotov}, \citenamefont {Narozhny}, \citenamefont {Mourou},\ and\
  \citenamefont {Korn}}]{FedotovPhRL2010}%
  \BibitemOpen
  \bibfield  {author} {\bibinfo {author} {\bibfnamefont {A.~M.}\ \bibnamefont
  {Fedotov}}, \bibinfo {author} {\bibfnamefont {N.~B.}\ \bibnamefont
  {Narozhny}}, \bibinfo {author} {\bibfnamefont {G.}~\bibnamefont {Mourou}}, \
  and\ \bibinfo {author} {\bibfnamefont {G.}~\bibnamefont {Korn}},\ }\href
  {\doibase 10.1103/PhysRevLett.105.080402} {\bibfield  {journal} {\bibinfo
  {journal} {Phys. Rev. Lett.}\ }\textbf {\bibinfo {volume} {105}},\ \bibinfo
  {pages} {080402} (\bibinfo {year} {2010})}\BibitemShut {NoStop}%
\bibitem [{\citenamefont {Bulanov}\ \emph {et~al.}(2010)\citenamefont
  {Bulanov}, \citenamefont {Esirkepov}, \citenamefont {Thomas}, \citenamefont
  {Koga},\ and\ \citenamefont {Bulanov}}]{BulanovPhRL2010}%
  \BibitemOpen
  \bibfield  {author} {\bibinfo {author} {\bibfnamefont {S.~S.}\ \bibnamefont
  {Bulanov}}, \bibinfo {author} {\bibfnamefont {T.~Z.}\ \bibnamefont
  {Esirkepov}}, \bibinfo {author} {\bibfnamefont {A.~G.~R.}\ \bibnamefont
  {Thomas}}, \bibinfo {author} {\bibfnamefont {J.~K.}\ \bibnamefont {Koga}}, \
  and\ \bibinfo {author} {\bibfnamefont {S.~V.}\ \bibnamefont {Bulanov}},\
  }\href {\doibase 10.1103/PhysRevLett.105.220407} {\bibfield  {journal}
  {\bibinfo  {journal} {Phys. Rev. Lett.}\ }\textbf {\bibinfo {volume} {105}},\
  \bibinfo {pages} {220407} (\bibinfo {year} {2010})}\BibitemShut {NoStop}%
\bibitem [{\citenamefont {Nerush}\ \emph {et~al.}(2011)\citenamefont {Nerush},
  \citenamefont {Kostyukov}, \citenamefont {Fedotov}, \citenamefont {Narozhny},
  \citenamefont {Elkina},\ and\ \citenamefont {Ruhl}}]{NerushPhR2011}%
  \BibitemOpen
  \bibfield  {author} {\bibinfo {author} {\bibfnamefont {E.~N.}\ \bibnamefont
  {Nerush}}, \bibinfo {author} {\bibfnamefont {I.~Y.}\ \bibnamefont
  {Kostyukov}}, \bibinfo {author} {\bibfnamefont {A.~M.}\ \bibnamefont
  {Fedotov}}, \bibinfo {author} {\bibfnamefont {N.~B.}\ \bibnamefont
  {Narozhny}}, \bibinfo {author} {\bibfnamefont {N.~V.}\ \bibnamefont
  {Elkina}}, \ and\ \bibinfo {author} {\bibfnamefont {H.}~\bibnamefont
  {Ruhl}},\ }\href {\doibase 10.1103/PhysRevLett.106.035001} {\bibfield
  {journal} {\bibinfo  {journal} {Phys. Rev. Lett.}\ }\textbf {\bibinfo
  {volume} {106}},\ \bibinfo {pages} {035001} (\bibinfo {year}
  {2011})}\BibitemShut {NoStop}%
\bibitem [{\citenamefont {Hu}\ \emph {et~al.}(2010)\citenamefont {Hu},
  \citenamefont {M\"uller},\ and\ \citenamefont {Keitel}}]{HuPhRL2010}%
  \BibitemOpen
  \bibfield  {author} {\bibinfo {author} {\bibfnamefont {H.}~\bibnamefont
  {Hu}}, \bibinfo {author} {\bibfnamefont {C.}~\bibnamefont {M\"uller}}, \ and\
  \bibinfo {author} {\bibfnamefont {C.~H.}\ \bibnamefont {Keitel}},\ }\href
  {\doibase 10.1103/PhysRevLett.105.080401} {\bibfield  {journal} {\bibinfo
  {journal} {Phys. Rev. Lett.}\ }\textbf {\bibinfo {volume} {105}},\ \bibinfo
  {pages} {080401} (\bibinfo {year} {2010})}\BibitemShut {NoStop}%
\bibitem [{\citenamefont {Lundstr\"om}\ \emph {et~al.}(2006)\citenamefont
  {Lundstr\"om}, \citenamefont {Brodin}, \citenamefont {Lundin}, \citenamefont
  {Marklund}, \citenamefont {Bingham}, \citenamefont {Collier}, \citenamefont
  {Mendon\ifmmode~\mbox{\c{c}}\else \c{c}\fi{}a},\ and\ \citenamefont
  {Norreys}}]{LundstromPhRL2006}%
  \BibitemOpen
  \bibfield  {author} {\bibinfo {author} {\bibfnamefont {E.}~\bibnamefont
  {Lundstr\"om}}, \bibinfo {author} {\bibfnamefont {G.}~\bibnamefont {Brodin}},
  \bibinfo {author} {\bibfnamefont {J.}~\bibnamefont {Lundin}}, \bibinfo
  {author} {\bibfnamefont {M.}~\bibnamefont {Marklund}}, \bibinfo {author}
  {\bibfnamefont {R.}~\bibnamefont {Bingham}}, \bibinfo {author} {\bibfnamefont
  {J.}~\bibnamefont {Collier}}, \bibinfo {author} {\bibfnamefont {J.~T.}\
  \bibnamefont {Mendon\ifmmode~\mbox{\c{c}}\else \c{c}\fi{}a}}, \ and\ \bibinfo
  {author} {\bibfnamefont {P.}~\bibnamefont {Norreys}},\ }\href {\doibase
  10.1103/PhysRevLett.96.083602} {\bibfield  {journal} {\bibinfo  {journal}
  {Phys. Rev. Lett.}\ }\textbf {\bibinfo {volume} {96}},\ \bibinfo {pages}
  {083602} (\bibinfo {year} {2006})}\BibitemShut {NoStop}%
\bibitem [{\citenamefont {Kuchiev}(2007)}]{KuchievPhRL2007}%
  \BibitemOpen
  \bibfield  {author} {\bibinfo {author} {\bibfnamefont {M.~Y.}\ \bibnamefont
  {Kuchiev}},\ }\href {\doibase 10.1103/PhysRevLett.99.130404} {\bibfield
  {journal} {\bibinfo  {journal} {Phys. Rev. Lett.}\ }\textbf {\bibinfo
  {volume} {99}},\ \bibinfo {pages} {130404} (\bibinfo {year}
  {2007})}\BibitemShut {NoStop}%
\bibitem [{\citenamefont {Kluger}\ \emph {et~al.}(1991)\citenamefont {Kluger},
  \citenamefont {Eisenberg}, \citenamefont {Svetitsky}, \citenamefont
  {Cooper},\ and\ \citenamefont {Mottola}}]{KlugerPhRL1991}%
  \BibitemOpen
  \bibfield  {author} {\bibinfo {author} {\bibfnamefont {Y.}~\bibnamefont
  {Kluger}}, \bibinfo {author} {\bibfnamefont {J.~M.}\ \bibnamefont
  {Eisenberg}}, \bibinfo {author} {\bibfnamefont {B.}~\bibnamefont
  {Svetitsky}}, \bibinfo {author} {\bibfnamefont {F.}~\bibnamefont {Cooper}}, \
  and\ \bibinfo {author} {\bibfnamefont {E.}~\bibnamefont {Mottola}},\ }\href
  {\doibase 10.1103/PhysRevLett.67.2427} {\bibfield  {journal} {\bibinfo
  {journal} {Phys. Rev. Lett.}\ }\textbf {\bibinfo {volume} {67}},\ \bibinfo
  {pages} {2427} (\bibinfo {year} {1991})}\BibitemShut {NoStop}%
\bibitem [{\citenamefont {Ruffini}\ \emph {et~al.}(2003)\citenamefont
  {Ruffini}, \citenamefont {Vitagliano},\ and\ \citenamefont
  {Xue}}]{RuffiniPLB2003}%
  \BibitemOpen
  \bibfield  {author} {\bibinfo {author} {\bibfnamefont {R.}~\bibnamefont
  {Ruffini}}, \bibinfo {author} {\bibfnamefont {L.}~\bibnamefont {Vitagliano}},
  \ and\ \bibinfo {author} {\bibfnamefont {S.-S.}\ \bibnamefont {Xue}},\ }\href
  {\doibase DOI: 10.1016/S0370-2693(03)00299-5} {\bibfield  {journal} {\bibinfo
   {journal} {Physics Letters B}\ }\textbf {\bibinfo {volume} {559}},\ \bibinfo
  {pages} {12 } (\bibinfo {year} {2003})}\BibitemShut {NoStop}%
\bibitem [{\citenamefont {Ruffini}\ \emph {et~al.}(2007)\citenamefont
  {Ruffini}, \citenamefont {Vereshchagin},\ and\ \citenamefont
  {Xue}}]{VereshchaginPLA2007}%
  \BibitemOpen
  \bibfield  {author} {\bibinfo {author} {\bibfnamefont {R.}~\bibnamefont
  {Ruffini}}, \bibinfo {author} {\bibfnamefont {G.}~\bibnamefont
  {Vereshchagin}}, \ and\ \bibinfo {author} {\bibfnamefont {S.-S.}\
  \bibnamefont {Xue}},\ }\href {\doibase DOI: 10.1016/j.physleta.2007.06.056}
  {\bibfield  {journal} {\bibinfo  {journal} {Physics Letters A}\ }\textbf
  {\bibinfo {volume} {371}},\ \bibinfo {pages} {399 } (\bibinfo {year}
  {2007})}\BibitemShut {NoStop}%
\bibitem [{\citenamefont {Benedetti}\ \emph {et~al.}(2011)\citenamefont
  {Benedetti}, \citenamefont {Han}, \citenamefont {Ruffini},\ and\
  \citenamefont {Vereshchagin}}]{Benedetti2011}%
  \BibitemOpen
  \bibfield  {author} {\bibinfo {author} {\bibfnamefont {A.}~\bibnamefont
  {Benedetti}}, \bibinfo {author} {\bibfnamefont {W.-B.}\ \bibnamefont {Han}},
  \bibinfo {author} {\bibfnamefont {R.}~\bibnamefont {Ruffini}}, \ and\
  \bibinfo {author} {\bibfnamefont {G.}~\bibnamefont {Vereshchagin}},\
  }\href@noop {} {\enquote {\bibinfo {title} {On the frequency of oscillations
  in the pair plasma generated by a strong electric field},}\ }\bibinfo
  {howpublished} {Physics Letters B, in press} (\bibinfo {year}
  {2011})\BibitemShut {NoStop}%
\bibitem [{\citenamefont {Berestetskii}\ \emph {et~al.}(1982)\citenamefont
  {Berestetskii}, \citenamefont {Lifshitz},\ and\ \citenamefont
  {Pitaevskii}}]{Landau_QED}%
  \BibitemOpen
  \bibfield  {author} {\bibinfo {author} {\bibfnamefont {V.~B.}\ \bibnamefont
  {Berestetskii}}, \bibinfo {author} {\bibfnamefont {E.~M.}\ \bibnamefont
  {Lifshitz}}, \ and\ \bibinfo {author} {\bibfnamefont {L.~P.}\ \bibnamefont
  {Pitaevskii}},\ }\href@noop {} {\emph {\bibinfo {title} {Quantum
  Electrodynamics}}}\ (\bibinfo  {publisher} {Pergamon Press},\ \bibinfo
  {address} {New York},\ \bibinfo {year} {1982})\BibitemShut {NoStop}%
\bibitem [{\citenamefont {Kokkotas}\ and\ \citenamefont
  {Schmidt}(1999)}]{KokkotasLivRev1999}%
  \BibitemOpen
  \bibfield  {author} {\bibinfo {author} {\bibfnamefont {K.~D.}\ \bibnamefont
  {Kokkotas}}\ and\ \bibinfo {author} {\bibfnamefont {B.~G.}\ \bibnamefont
  {Schmidt}},\ }\href@noop {} {\bibfield  {journal} {\bibinfo  {journal}
  {Living Reviews in Relativity}\ }\textbf {\bibinfo {volume} {2}},\ \bibinfo
  {pages} {2} (\bibinfo {year} {1999})}\BibitemShut {NoStop}%
\end{thebibliography}%

\end{document}